\documentclass[twocolumn,twoside,slac_two]{revtex4}
\usepackage{graphicx}
\usepackage{fancyhdr}
\usepackage{empheq}
\usepackage{wasysym}
\usepackage{pstricks}
\usepackage{color}
\usepackage{bbm}
\def\qq{\mathbbmss{q}}
\def\qh{\mathbbmss{Q}}

\begin{document}

\title{Hints of a New Spectroscopy}

%

\author{AD Polosa}
\affiliation{INFN Roma, Piazzale A Moro 2, Roma, I-00185, Italy}

\begin{abstract}
There are several reasons to believe that some of the new particles observed at $B$-factories 
have no-ordinary quark composition. We briefly illustrate the diquark-antidiquark 
model and the recent experimental discoveries which confirm some of its most striking predictions.

\end{abstract}

\maketitle

\thispagestyle{fancy}


\section{Introduction}
The field of hadron spectroscopy has been revitalized by the discovery of a number of new particles at $B$-factories. The first of the series is a $1^{++}$ state, the $X(3872)$, decaying to $J/\psi \pi\pi$, found by Belle and later confirmed by  CDF, D0 and BaBar~\cite{alldata}. The difficulty at interpreting this particle as a standard charmonium has opened the way to alternative interpretations. The diquark-antidiquark, $[cq][\bar c \bar q]$ picture~\cite{mppr}, at the moment, seems the most promising. There are at least two reasons for this:
\begin{itemize}
\item The $X(3872)$ decays with the same rate in $J/\psi \rho$ and $J/\psi \omega$, therefore maximally violating isospin. In the diquark picture, two states are needed to explain this decay pattern, namely an $X_u$ and an $X_d$, with a difference in mass of the order of  $m_u-m_d$. Recently Belle and BaBar have shown the existence of a second $X$ at a mass of $3875$~MeV, confirming the diquark model prediction~\cite{newX}.
\item The stringiest prediction of the diquark model was the existence of charged  particles decaying to $J/\psi (\pi^\pm \vee \rho^\pm)$~\cite{mppr}. Last summer a state of this kind has been discovered by Belle: the $Z(4430)$ decaying to $J/\psi \pi^+$~\cite{newZ}.
\end{itemize}

After the discovery of the $X(3872)$, BaBar has found a new state produced in ISR, the $Y(4260)$~\cite{y4260}. Again a charmonium interpretation was not tenable for this particle, despite its decay to $J/\psi \pi\pi$. The diquark model  explains the $Y(4260)$ as a $1^{--}$ state made up of two diquarks in P-wave which decays to $f_0(980)$, a strong candidate for a light four-quark state, and $J/\psi$. The increasing number of $X,Y,Z$ states being found, casts serious doubts on the fact that all of them can be explained, separately, as effects (threshold effects, cusp effects...) or loosely bound molecules~\cite{others} of open charm mesons. If other ways of aggregation of quarks in matter are indeed possible, a unified explanation of all these particles, and of the coming ones (hopefully!), will emerge clearly from data.

\section{From Light Scalars to $XYZ$}
The strongest theoretical motivation for the diquark-antidiquark picture lies in its consistent description of the scalar mesons below $1$~GeV, namely $f_0, a_0,\kappa,\sigma$. 
These are likely bound states of a spin zero diquark
\begin{equation}
\qq_{i\alpha}= \epsilon_{ijk}\epsilon_{\alpha \beta\gamma} \bar q^{j\beta}_C\gamma_5 q^{k\gamma},
\label{defdq}
\end{equation}
where latin indices label flavor and greek letters are for color, and an anti-diquark ${\bar \qq}^{i\alpha}$. The color is saturated as in a standard $q\bar q$ meson: $\qq^\alpha {\bar \qq}_\alpha$. Therefore, since a spin zero diquark is in a ${\bf \bar 3}$-flavor representation,  nonets of $\qq\bar \qq$ states are allowed (crypto-exotic states). The sub-GeV scalar mesons represent most likely in the lowest tetraquark nonet. 

The $\qq\bar \qq$ model of light-scalars is very effective at explaining the most striking feature of these particles: the inverted pattern in the mass-versus-$I_3$ diagram~\cite{jaffe}, with respect to what observed for ordinary $q\bar q$ mesons.This aspect is not explicable using a $q\bar q$ model. 
For example, in the $q\bar q$ model, the $f_0(980)$ should be an $s\bar s$ state~\cite{noitornq} while the
$I=1$, $a_0(980)$, should be a $u\bar u+d\bar d$ state. If this were the case, the degeneracy of the two particles would appear quite mysterious. 

Beyond a correct description of the mass-$I_3$ pattern, the tetraquark model offers the possibility to explain the decay rates of scalars at a level never reached in standard $q\bar q$ descriptions.
The decay Lagrangian into two pseudoscalar mesons, e.g. $\sigma \to \pi\pi$, is:
\begin{equation}
{\cal L}_{\rm exch.}=c_fS^i_j \epsilon^{j t u}\epsilon_{i r s} \Pi^r_t\Pi^s_u,
\label{exchange}
\end{equation}
where $i,j$ are the flavor labels of $\qq^i$ and $\bar\qq^j$, while $r,s,t,u$ are the flavor labels of the quarks $\bar q^t,\bar q^u$ and  $ q^r, q^s$. $c_f$ is the effective coupling weighting this interaction term and ${\bf S},{\bf \Pi}$ are the scalar and pseudoscalar matrices. This Lagrangian describes the quark exchange amplitude  for  the quarks to tunnel out of their diquark shells to form ordinary mesons~\cite{newlook}. Such mechanism is the alternative to the color string breaking  $\qq\gluon q\bar q\gluon\bar \qq\to B\bar B$, {\it i.e.}, a baryon-anti-baryon decay, phase-space forbidden to sub-GeV scalar mesons.

The problem with~(\ref{exchange}) is that it is not able to describe the decay $f_0\to \pi\pi$ because $f_0=( \qq^2{\bar\qq}^2+\qq^1\bar \qq^1)/\sqrt{2}$, being $1,2,3$ the $u,d,s$ flavors so that, see equation~(\ref{defdq}),  $\qq^1=[ds]$ and $\qq^2=[su]$. An annihilation diagram is needed to replace the $s$ quarks. This clearly induces a small rate which does not match the observed one. 

We have faced this problem recently and we are confident to have found a solid solution which generally improves the agreement with data  of all light scalar mesons decay rates~\cite{winprog}.

\section{Heavy-Light Diquarks}
The  successful theoretical interpretation of the light scalar mesons in terms of diquarks suggest that such structures could exist also at higher mass scales. An heavy-light diquark is still bound in color ${\bf \bar 3}_c$, but: {\bf 1.} We cannot state anymore that there is a Fermi statistics that forces the diquarks to be in ${\bf\bar 3}_f$ or ${\bf 6}_f$ depending if the diquark spin is zero or one (a charm quark by no way can be considered identical to a light $q$ with respect to strong interactions); {\bf 2.} The spin-spin interaction is weakened by $1/m_c$. According to lattice studies, light diquarks are preferably formed with spin$=0$. On the other hand, heavy-light diquarks could appear equally well in  spin $0$ and
$1$. Therefore our {\it Ansatz} for  heavy-light diquarks is:
\begin{eqnarray}
\qh_{1\alpha}^i=\epsilon_{\alpha\beta\gamma}\bar Q^{\beta}_C\gamma_5\vec{\gamma} q^{i\gamma}&&{\rm spin}~1^-\\
\qh_{0\alpha}^i=\epsilon_{\alpha\beta\gamma}\bar Q^{\beta}_C\gamma_5 q^{i\gamma}&&{\rm spin}~0^+
\end{eqnarray}
The flavor $i$ is carried by the light quark, while $Q=c$ for all $X,Y,Z$. Such spin 1 diquarks are likely the building blocks of the new particles, and since the flavor of $\qh$ is  the flavor of the light quark, we can still accommodate particles like the $X(3872)$ and its partners in $SU(3)$ nonets~\cite{mppr}. In our notation:
\begin{eqnarray}
&&X(3872)=X_d=\qh^2_1\bar \qh^2_0+\qh^2_0\bar \qh^2_1\\
&&X(3875)=X_u=\qh^1_1\bar \qh^1_0+\qh^1_0\bar \qh^1_1\\
&&Y(4260)=(\qh^3_0\bar \qh^3_0)_{\rm P-wave}\\
&&Z(4433)=(\qh^1_1\bar\qh^2_0+\qh^1_0\bar\qh^2_1)_{\rm 2S-wave}.
\end{eqnarray}

We have shown in~\cite{newprl} that, assuming:
\begin{eqnarray}
&&X_d\to J/\psi \pi^+\pi^-\\
&&X_u\to D^0\bar D^0\pi^0,
\end{eqnarray}
we obtain a simple rule for the ratios of branching ratios of $B$ decays. With an obvious notation:
\begin{equation}
\left.\frac{B^0}{B^+}\right|_{K J/\psi \pi\pi}= \left(\left.\frac{B^0}{B^+}\right|_{K D\bar D \pi}\right)^{-1}.
\end{equation}
This is to be confronted with the most recent experimental data, giving:
\begin{equation}
0.94\pm 0.24\pm 0.10=\frac{1}{1.33\pm 0.69\pm 0.52}
\end{equation}

Such quite reasonable agreement gives credit to our assignations for $X(3872)$ and $X(3875)$.
Clearly enough, the situation of the remaining $1^{++}$ {\it charged} partners has to be clarifed. We have reasons to believe that they can be very broad and this could  cast some doubts on their actual visibility.  
\begin{table}[h]
\begin{center}
\caption{Some $L=0$ neutral tetraquark wave functions constructible from diquarks. Consider that
$C|f\bar f\rangle=(-1)^{L+S}|f\bar f\rangle$.}
\begin{tabular}{|c|c|}
\hline \textbf{$J^{PC}$} & \textbf{Wave Funct.} 
\\
\hline $0^{++}$ & $\qh_0\bar \qh_0 \vee (\qh_1\bar \qh_1)_{J=0} $\\
\hline $1^{++}$ & $\frac{\qh_1\bar \qh_0+\qh_0\bar \qh_1}{\sqrt{2}} $\\
\hline $1^{+-}$ & $\frac{\qh_1\bar \qh_0-\qh_0\bar \qh_1}{\sqrt{2}}\vee  (\qh_1\bar \qh_1)_{J=1} $\\
\hline $2^{++}$ & $ (\qh_1\bar \qh_1)_{J=2} $\\ 
\hline
\end{tabular}
\label{wavefunct}
\end{center}
\end{table}

\section{$Z(4433)$}
As mentioned in the introduction, the $Z(4433)$, most likely a $1^{+-}$ state, is the first charged particle observed that plausibly fits very well  a diquark-antidiquark interpretation. Due to its decay to $\psi(2S)\pi^+$  we think that $Z(4433)$ is itself a radial excitation of one of the lowest lying $1^{+-}$ states predicted by the tetraquark model  to be at a mass of $\sim 3880$~MeV~\cite{mppr}. In particular it is striking to observe that:
\begin{equation}
M(\psi(2S))-M(J/\psi)\simeq 590~{\rm MeV}\simeq 4433-3880~{\rm MeV}\notag
\end{equation}
So the search of $X(1^{+-};1S)$ states, neutral and charged, is particularly urgent. 

Scanning higher energy regions one approaches  baryon-baryon thresholds, like $2M_{\Lambda_c}=4572$~MeV and $M_{\Lambda_c}+M_{\Sigma_c}=4379$~MeV. As mentioned above, the baryon-antibaryon decay channel is expected to be quite natural for a diquark-antidiquark system. This would mean that, above a certain threshold, all charmed tetraquark states are expected to  be very broad.

\section{1/N and Tetraquarks}
In the large number of colors $1/N$ expansion, four-quark states of the kind $\bar q_\alpha q^\alpha \bar q_\beta q^\beta(={\cal O}(x)) $
are suppressed, in the sense that they do not appear as  leading terms in the $1/N$ expansion of the 
two-point correlation function $\langle {\cal O}(x){\cal O}(0)\rangle$. Indeed, the leading term of such a correlator would represent a disconnected graph (see the left diagram in Fig.~1).
\begin{figure}[h]
\hspace{-2.0truecm}
\includegraphics[width=80mm]{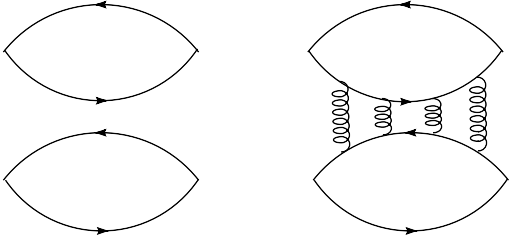}
\caption{Color diagrams for a tetraquark $\bar q_\alpha q^\alpha \bar q_\beta q^\beta$. The disconnected diagram is
$O(N^2)$ larger than the connected one (bounded meson) in the `t Hooft expansion~\cite{thooft}.} \label{example_figure}
\end{figure}

But if we replace quarks with diquarks: $q^\alpha\to\qq^\alpha$, there are no disconnected parts and the leading color diagram looks like the one depicted in Fig.~2.
\begin{figure}[h]
\hspace{-1.0truecm}
\includegraphics[width=30mm]{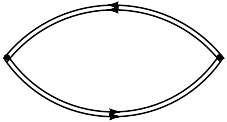}
\caption{Color diagram for a tetraquark $\bar \qq_\alpha\qq^\alpha$}
\end{figure}
In other words, if the diquarks are assumed to be the building blocks, we evade the $1/N$ difficulty with standard () tetraquarks.

This reminds of the Corrigan-Ramond (CR)~\cite{CR} large $N$ limit where quarks and  `larks' are introduced, transforming as the ${\bf N}$ and ${\bf N(N-1)/2}$ representations of $SU(N)$. Larks, $\ell$, are therefore antisymmetric objects,  $\ell_{\alpha\beta}=-\ell_{\beta\alpha}$, coinciding with antiquarks if $N=3$. A theory of only larks is equivalent to QCD. In the CR expansion, it is therefore possible to consider baryons at large $N$ , if the baryon is represented by a color saturated $qq\bar \ell$ state (for three quarks one can neutralize the color with an $\epsilon_{\alpha\beta\gamma}$ if the colors are three: there are no color singlets made up of  three quarks for $N>3$). Since larks have two color indices (like gluons in the `t Hooft limit~\cite{thooft}, but with the arrow pointing in the same directions - gluon lines $A^\alpha_\beta$ have opposite arrows - ), the large $N$ power counting works differently (see Fig. 3). A diquark-antidiquark state is then like a lark-antilark state $\ell\bar \ell$ (see Fig. 2).

\begin{figure}[h]
\includegraphics[width=60mm]{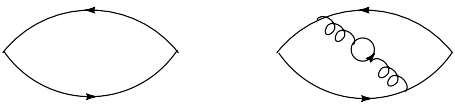}
\caption{In the CR expansion, the $\ell\bar \ell$ two point correlation function on the left and the correlator on the right (where a four-lark structure emerge in the intermediate state), are of the same $1/N$ order. In the `t Hooft limit, and with only quarks, the left diagram dominates, being larger by $O(N)$. } \label{example_figure}
\end{figure}

\section{Conclusions: Beyond Spectroscopy}
The interest for such problems has implications beyond spectroscopy itself. This is not merely a chemistry exercise, aimed at a classification of new `elements'  whose nature has very little impact to QCD fundamental issues. QCD is extremely predictive only in a narrow range of very high energy phenomena, while the study of hadron structures and interactions remains a very difficult field which is mainly approached with the use of Effective Theories and Lattice studies. Being able to assess that new forms of aggregation of quarks are possible, such as diquarks, opens a window on a territory poorly known. 
Moreover, diquarks are essential to the theory of color superconductivity~\cite{raj}, which is at the basis of the comprehension of  an entirely new sector of the QCD phase diagram.

The skepticism of the community about the tetraquarks is mainly driven by the shock following the `discovery' and  the disappearance of the baryonic pentaquarks. It is quite possible that multiquark baryons might exist at higher masses. Anyway, there is no clear logical connection between the four-quark mesons we are discussing here and the pentaquark baryons; one should also remind that the case for  four-quark mesons is based on the phenomenology of light scalars since a very long time.
Some other recent investigations on diquark based tetraquark charmed mesons can be found in~\cite{oth4q}.

We believe that a strong experimental effort aimed at searching the new predicted particles and clearly discriminate between models is of great importance for a progress in the comprehension of key aspects of non-perturbative QCD.

\begin{acknowledgments}
I acknowledge the Physics Department of Roma `La Sapienza'.
I wish to thank Luciano Maiani for many discussions and fruitful collaboration.
\end{acknowledgments}

\bigskip 

\end{document}